\documentstyle[seceq]{ptptex}

\notypesetlogo  

\title{Quantum Analysis and Nonequilibrium Response        
}

\author{Masuo {\sc Suzuki}       
}

\inst{Department of Applied Physics, Science University of Tokyo\\
Tokyo 162-8601, Japan          
}

\recdate{April 3, 1998      
}
\abst{
The quantum derivatives of $e^{-A}, A^{-1}$ and $\log A$, 
which play a basic role in quantum statistical physics, are derived 
and their convergence is proven for an unbounded positive 
operator $A$ in a Hilbert space. Using the quantum analysis
based on these quantum derivatives,
a basic equation for the entropy operator in
nonequilibrium systems is derived, and 
Zubarev's theory is extended to 
infinite order with respect to a perturbation.
Using the first-order term of this general 
perturbational expansion of the entropy operator,
Kubo's linear response is rederived and expressed
in terms of 
the inner derivation $\delta_{{\cal H}}$ for the relevant 
Hamiltonian ${\cal H}$.   Some 
remarks on the conductivity $\sigma (\omega )$ are given. 
}

\begin{document}

\maketitle

\section{Introduction}
Recently, the present author$^{1)-3)}$ proposed a new scheme of 
quantum calculus, the so-called quantum analysis. In this scheme, the 
derivative of an operator-valued function with respect to the 
relevant operator itself is expressed only in terms of the original 
operator and its inner derivation (i.e., a hyperoperator or 
superoperator), and an operator expansion formula is derived. 

In the present paper, the quantum derivatives of $e^{-A}, A^{-1}$
and $\log A$, which are basic operator functions in physics, are derived,
and their convergence is proven in \S~2 for the unbounded positive 
operator $A$ in a Hilbert space. (See also Appendices A and B.)
Nonlinear responses in equilibrium
are expressed in terms of
quantum derivatives 
in \S~3.  A basic equation
for nonequilibrium systems is derived in \S~4 using quantum
analysis. 
This derivation has the merit that it is valid even for an unbounded
entropy operator. On the other hand, Zubarev's derivation is based on the
power series expansion of the density matrix with respect to
the entropy operator, and consequently it is restricted to a bounded 
entropy operator.
Zubarev's theory$^{4)}$ is extended  
to infinite order in \S~5. This gives a renormalized perturbation theory with 
respect to an external field. 
Kubo's formula of linear response$^{5),6)}$ is then rederived and 
expressed in terms of an inner derivation in \S~6.  
Some remarks on the conductivity $\sigma(\omega)$ are given in \S~7. 
The entropy operator $\eta(t)$ in a 
dissipative system [namely $-\log \rho(t)$ for the 
density matrix 
$\rho(t)$] is expressed in a compact form using the inner derivation in Appendix C. 
This expression is convenient for studying quantum effects, because it 
is expressed only in terms of commutators.  

\section{\bf Quantum derivatives of ${\rm e}^{-A}$,     
$A^{-1}$ and $\log A$
and the convergence of the differential $df(A)$}

In a previous paper,$^{1)}$ quantum analysis was formulated in a 
Banach space, namely for bounded operators. The term `quantum analysis'  
refers to noncommutative differential calculus in terms of inner 
derivations, namely commutators. Formal expressions and several 
formulas of quantum analysis are derived in Ref.~1). In 
practical applications, for example, to quantum and 
statistical physics, we often have to treat unbounded operators in a Hilbert space. 
As is well known, it is difficult to prove generally the convergence of 
such formal expressions for unbounded operators.$^{7),8)}$ 
Fortunately, the density matrix $\rho$ in statistical mechanics is a 
contraction operator when the relevant Hamiltonian ${\cal H}$ is 
unbounded (even for a finite system) but positive definite (or
bounded below). 
Furthermore, a perturbation may often be assumed to be bounded in 
statistical physics. 
(For example, a Zeeman energy is expressed by a bounded
operator in a finite system, while the kinetic energy of
an itinerant electron system is unbounded.)
Hereafter we discuss the quantum calculus for 
these situations. Thus we study here the convergence of the  
G\^ateaux differential                       
\begin{equation}
df(A)=\lim_{h \rightarrow 0} \frac{f(A+hdA)-f(A)}{h},  
\end{equation}
where $A$ is unbounded but $f(A)$ is bounded,
and $dA$ is an arbitrary bounded operator independent of $A$. To consider 
this situation is one of the key points for studying the convergence 
of Eq.~(2$\cdot $1).  The quantum derivative $df(A)/dA$ is defined$^{1)}$ by

\begin{subequations}
 \label{eq :1}

\begin{equation}
df(A) = \frac{df(A)}{dA} dA. 
 \label{eq :1a}  
\end{equation}
Here $df(A)/dA$ is a hyperoperator which is a function of both $A$ and the inner
derivation $\delta_A$ defined by Eq.~(2$\cdot $7b).
This property is crutial in quantum analysis.$^{1)}$
In fact, we have the formula$^{1)-3)}$
\begin{equation}
\frac{df(A)}{dA} = \frac{\delta_{f(A)}}{\delta_A} 
 \\
\end{equation}
\end{subequations}
in a Banach space.  Higher-order quantum derivatives will
be discussed in \S~3.

\vspace*{0.3cm}
\noindent
i) {\it Quntum derivatives of \, ${\rm e}^{-A}$ and $A^{-1}$}

Here, we attempt to prove the convergence of Eq.~(2$\cdot$1) for two typical 
operator functions, $f(A)={\rm e}^{-A}$ and $f(A)=A^{-1}$, where $A$ is a 
positive (but unbounded) operator. Clearly we have 

\begin{equation}
\mid \mid {\rm e}^{-A} \mid \mid <1 \; \; {\rm and} \; \; \mid \mid 
A^{-1} \mid \mid <\infty,  
\end{equation}
under the condition that $A \geq a>0$ for a constant $a$.

First note that$^{7),9)}$

\begin{subequations}
 \label{eq :1}

\begin{equation}
\frac{d}{dx} {\rm e}^{-(A+xB)}
=-\int^1_0 {\rm e}^{-(1-s)(A+xB)} B {\rm e}^{-s(A+xB)} ds. 
 \label{eq :1a}  
\end{equation}
Integrating Eq.~(2$\cdot$4a) we obtain
\begin{equation}
{\rm e}^{-(A+xB)}
={\rm e}^{-A}
-\int^x_0 dt \int^1_0 ds {\rm e}^{-(1-s)(A+tB)} B {\rm e}^{-s(A+tB)}. 
\end{equation}

\end{subequations}

Then we can prove the convergence
\begin{equation}
\lim_{h \rightarrow 0} \mid \mid
({\rm e}^{-A+hB)}-{\rm e}^{-A})/h
-\int^1_0 {\rm e}^{-(1-s)A}(-B) {\rm e}^{-sA} ds \mid
\mid = 0
\end{equation}
when $A$ is a positive (but unbounded) operator and $B = dA$
is bounded, as is shown in detail in Appendix A.
Thus we arrive at the differential

\begin{subequations}
 \label{eq :1}

\begin{equation}
d({\rm e}^{-A}) = -\int^1_0 {\rm e}^{-(1-s)A} (dA){\rm e}^{-sA}ds 
=-{\rm e}^{-A}\Delta (A) dA,
 \label{eq :1a}  
\end{equation}
or the quantum derivative
\begin{equation}
\frac{d{\rm e}^{-A}}{dA}=-{\rm e}^{-A}+
\int^1_0 {\rm e}^{-(1-s)A}\delta_{{\rm exp}(-sA)}ds
= -{\rm e}^{-A} \Delta (A).    
\end{equation}
\end{subequations}
This is well defined for a positive operator $A$. Here, the hyperoperator
$\Delta(A)$ is defined by$^{1)}$

\begin{subequations}
 \label{eq :1}

\begin{equation}
\Delta(A) = \int^1_0 {\rm e}^{t\delta_A} dt=\frac{{\rm 
e}^{\delta_A}-1}{\delta_A}, 
 \label{eq :1a}  
\end{equation}
with the inner derivation $\delta_A$ defined by
\begin{equation}
\delta_A Q \equiv [A,Q] \equiv AQ-QA.
 \end{equation}
\end{subequations}
The ratio of the hyperoperators $({\rm e}^{\delta_A}-1)$ and $\delta_A$ is 
well defined, although $\delta_A^{-1}$ does not necessarily exist.
The formula (2$\cdot$6b) with Eq.~(2$\cdot$7a) will be used frequently later. 

Concerning the convergence of the power series expansion of 
${\rm e}^{-A} \Delta (xA)$, we have the theorem. 

\vspace*{0.3cm}
\noindent
{\bf Theorem 1 :} {\it The power series expansion of} ${\rm 
e}^{-A}\Delta(xA)dA$ {\it with respect to x converges in the uniform norm 
topology for} $A>0$ {\it and for} $\mid 
x \mid < \alpha^{-1}$, {\it where} $\alpha $ {\it is defined by the 
upper limit} 
\begin{equation}
\alpha=\lim^{\mbox{---}}_{n \rightarrow \infty} \mid \mid (A^{-1} 
\delta_A)^n dA \mid \mid^{\frac{1}{n}}. 
\end{equation}

The proof is easily given using the Stirling formula $n ! \simeq n^n 
{\rm e}^{-n}$ for large $n$ and the following inequality. 

\vspace*{0.3cm}
\noindent
{\bf Inequality} : When $A>0$, we have

\begin{subequations}
 \label{eq :1}

\begin{equation}
\mid \mid {\rm e}^{-A}\delta^n_A B \mid \mid \,\, 
\leq \,\, n^n {\rm e}^{-n}  \mid \mid (A^{-1} \delta_A)^n B 
\mid \mid 
 \label{eq :1a}  
\end{equation}
for any positive integer $n$.

The proof of the above inequality is 
easily given using the inequalities
\begin{equation}
\mid \mid {\rm e}^{-A} \delta^n_A B \mid \mid \,\, 
\leq \,\, \mid \mid {\rm e}^{-A} A^n \mid \mid \cdot 
\mid \mid (A^{-1} \delta_A)^n B \mid \mid  
\end{equation}
and
\begin{equation}
\mid \mid {\rm e}^{-A} A^n \mid \mid \,\,
\leq \,\, {\rm e}^{-n} n^n.   
\end{equation}

\end{subequations}

It should be noted here that $\delta_A$ and $A$ 
commute.

\vspace*{0.3cm}
\noindent
{\bf Corollary} : If $B^{1/k}$ is defined for any positive integer $k$
and there exists the maximum $M \equiv {\rm max}_k \mid \mid A^{-1} 
B^{1/k} A \mid \mid $, then the power series expansion of 
${\rm e}^{-A} \Delta (xA) B $ with respect to $x$ converges
in the uniform norm topology for $ A > 0 $ and for
$ \mid x \mid < 1/(M + 1) $.

Proof: First note that

\begin{subequations}
 \label{eq :1}

\begin{equation}
(A^{-1} \delta_A)^n B = \sum_{k=0}^n (-1)^k {n\choose k}
A^{-k} B A^k 
 \label{eq :1a}  
\end{equation}
for any positive integer $n$. Then we have

\begin{eqnarray}
\mid \mid (A^{-1} \delta_A)^n B \mid \mid   \leq 
\sum_{k=0}^n {n\choose k} \mid \mid
A^{-k} B A^k \mid \mid~~~~~~  \nonumber \\
  \leq \sum_{k=0}^n {n\choose k} 
\mid \mid A^{-1} B^{1/k} A \mid \mid ^k
\leq (M + 1)^n
\end{eqnarray}
under the conditions of the above corollary. Thus we obtain
\begin{equation}
{\lim^{\mbox{---}}_{n \rightarrow \infty}} \mid \mid
(A^{-1} \delta_A)^n B \mid \mid ^{
\frac{1}{n}} ~\leq~ M + 1.  
\end{equation}
\end{subequations}
Note that~ 
${\rm lim}_{k \rightarrow \infty} \mid \mid
A^{-1} B^{1/k} A \mid \mid = 1$. Then the maximum number
$M$ may exist when the deformation of
$B^{1/k}$ by the transformation of an unbounded operator
$A$ is finite for all values of {\it k}. 
 
Similarly we study the differential of the resolvent operator $A^{-1}$  when 
$A \geq a >0$. We easily obtain

\begin{subequations}
 \label{eq :1}

\begin{eqnarray}
\lim_{h \rightarrow 0}
\mid \mid 
\left(\frac{1}{A+hB}-\frac{1}{A}\right)/h-\frac{1}{A}(-B)\frac{1}{A} 
\mid \mid~~~~~~~~~~\nonumber \\ 
\leq \lim_{h \rightarrow 0} \mid h \mid 
\cdot \mid \mid \frac{1}{A} \mid \mid^2 \cdot \mid \mid 
\frac{1}{A+hB} \mid \mid \cdot \mid \mid B \mid 
\mid^2=0.  
 \label{eq :1a}   
\end{eqnarray}
That is, we have
\begin{equation}
d(\frac{1}{A})=-\frac{1}{A}(dA) \frac{1}{A}= (-A^{-2} +A^{-1} 
\delta_{A^{-1}}) dA=
-\frac{1}{A(A-\delta_A)}dA.   
\end{equation}
This gives
\begin{equation}
\frac{d}{dA} (\frac{1}{A}) = - \frac{1}{A(A-\delta_A)}.   
\end{equation}
\end{subequations}
This is also bounded when $A \geq a > 0$. Here we have used 
the relation$^{1)}$ $\delta_{A^{-1}} = A^{-1} - (A - \delta_{A})^{-1}$. 

\vspace*{0.3cm}
\noindent
ii) {\it Quantum derivative of $\log A$}

The above arguments can be extended to more general case in which 
$f(A)$ is also unbounded but $df(A)$ is bounded for bounded $B=dA$. A 
typical case is given by $f(A)=\log A$. The operator 
$\log(A+hB)$ is formally expressed by the following integral 
\begin{eqnarray}
\log(A+hB)&=&\int^{\infty}_0\left(\frac{1}{t+1}-\frac{1}{t+A+hB}\right)dt\nonumber\\ 
& = & \int^{\infty}_0\left(\frac{1}{t+1}-\frac{1}{t+A}\right)dt +h 
\int^{\infty}_0 \frac{1}{t+A}B \frac{1}{t+A}dt\nonumber\\ & & -h^2 
\int^{\infty}_0 \frac{1}{t+A}B \frac{1}{t+A}B 
\frac{1}{t+A+hB}dt. 
\end{eqnarray}
Then we have
\begin{eqnarray}
\lefteqn{\mid \mid [\log(A+hB)-\log A]/h - 
\int^{\infty}_0\frac{1}{t+A}B \frac{1}{t+A}dt\mid \mid}\nonumber\\ & 
& \leq \mid h \mid \cdot \mid \mid B \mid \mid^2 \int^{\infty}_0 \mid 
\mid \frac{1}{t+A} \mid \mid^2 \cdot \mid \mid \frac{1}{t+A+hB} \mid 
\mid dt.   
\end{eqnarray}
Consequently we arrive at$^{10)}$
\begin{equation}
d \log A=
\int^{\infty}_0 \frac{1}{t+A} (dA) \frac{1}{t+A}dt. 
\end{equation}
Clearly this is bounded when $A$ is positive (i.e., $A \geq a >0$) 
and $dA$ is bounded. This is formally written as
\begin{equation}
\frac{d \log A}{dA}=\frac{1}{A} -\int^{\infty}_0 \frac{1}{t+A} 
\delta_{(t+A)^{-1}}dt =
-\delta^{-1}_{A} \log(1-A^{-1} \delta_A). 
\end{equation}
The second expression of Eq.~(2$\cdot$15) gives the convergence of $df(A)/dA$.

\vspace{0.3cm}
\noindent
iii) {\it Convergence of $df(A)$ for an unbounded operator $A$
and for the bounded differential $dA$.}

 In general, the derivative
$df(A)/dA$ is formally given by the following formula.$^{1)}$

\vspace*{0.3cm}
\noindent
{\bf Formula 1 :} When $f(x)$ is an analytic function of $x$, we 
have
\begin{equation}
\frac{df(A)}{dA}=\frac{\delta_{f(A)}}{\delta_A}= 
\frac{f(A)-f(A-\delta_A)}{\delta_A} =
\int^1_0f^{(1)}(A-t \delta_A)dt. 
\end{equation}
Here $f^{(n)}(x)$ denotes the $n$th derivative of $f(x)$. This is 
formally expanded as
\begin{equation}
\frac{df(A)}{dA}=f^{(1)}(A)-\frac{1}{2!} f^{(2)}(A) \delta_A+ \cdots 
+ \frac{(-1)^n}{(n+1)!} f^{(n+1)} (A) \delta^n_A + \cdots. 
\end{equation}

Then, we have the following theorem.

\vspace*{0.3cm}
\noindent
{\bf Theorem 2 :} {\it Let} $A$ {\it be unbounded, and let} 
$\{f^{(n)}(A)\}$ {\it for} $n=0,1,2, \cdots$ {\it and} $dA$ {\it be 
bounded}. {\it Then, the formal expansion (2$\cdot $17) operating on} $dA$ 
{\it converges to Eq.~(2$\cdot$16) in the uniform norm topology if}

\begin{subequations}
 \label{eq :1}

\begin{equation}
\alpha_{1} \equiv \lim^{\mbox{---}}_{n \rightarrow \infty} \mid \mid 
\frac{f^{(n+1)}(A)}{(n+1)!}\delta^n_A dA \mid \mid^{\frac{1}{n}} <1. 
 \label{eq :1a}   
\end{equation}

A proof of this theorem is easily obtained. Theorem 1 is a typical 
example of the above general theorem. This theorem can also be 
extended to higher-order derivatives (see \S~3 and Appendix B). 

In the more general situation in which the operators $\{f^{(n+1)}(A) 
\delta^n_A dA\}$ are unbounded, the convergence proof of Eq.~(2$\cdot$17) can be 
studied using the strong norm convergence. Then, the condition (2$\cdot$18a) 
is replaced by 
\begin{equation}
\alpha'_{1} \equiv \lim^{\mbox{---}}_{n \rightarrow \infty} \mid \mid 
\frac{f^{(n+1)}(A)}{(n+1)!}\delta^n_A dA \psi \mid \mid^{\frac{1}{n}} 
<1      
\end{equation}
\end{subequations}
for $\psi \in {\cal D}$ with some appropriate domain ${\cal D}$ in  
Hilbert space.

\section{Higher-order quantum derivatives and
nonlinear responses in equilibrium}

The higher-order quantum derivative $d^nf(A)/d^nA$
is formally expressed $^{1)}$ by the multiple integral
\begin{equation}
\frac{d^nf(A)}{dA^n} = n! \int^1_0 dt_1 \int^{t_1}_0 dt_2
\cdots \int^{t_{n-1}}_0 dt_n \, f^{(n)} (A-t_1 \delta_1 - 
\cdots -t_n \delta_n),
\end{equation}
where $f(x)$ denotes the {\it n}-th derivative of $f(x)$
and the inner derivation $\delta_j$ is defined by
\begin{equation}
\delta_j : dA \cdot dA \cdot \cdots \cdot dA
= dA \cdot dA \cdot \cdots \cdot (\delta_A dA)
\cdot \cdots \cdot dA.
\end{equation}
Then we have the following operator Taylor expansion 
formula :$^{1),2)}$
\begin{equation}
f(A + xB) = \sum^{\infty}_{n=0} \frac{x^n}{n!} \,\,
\frac{d^n f(A)}{dA^n} : B^n
\end{equation}
with the notation $B^n = B \cdot \cdots \cdot B$ .

It is sometimes important to study nonlinear responses in condensed 
matter physics, as in spin glasses (in which only nonlinear 
susceptibilities diverge$^{11),12)}$ at the transition point). 

As is well known, an equilibrium system is described by the 
canonical density matrix 
\begin{equation}
\rho={\rm e}^{-\beta({\cal H}-HQ)} 
\end{equation}
for the Hamiltonian ${\cal H}$ of the system in the presence of an 
external field $H$ conjugate to a physical quantity $Q$. When $Q$ 
does not commute with ${\cal H}$, nonlinear responses are described 
in terms of the canonical correlations of $Q$, namely 
by a multiple integral of the time correlation function of $Q$
using the Feynman formula.
They are now expressed as 
\begin{equation}
\rho=\sum^{\infty}_{n=0}\frac{(-H)^n}{n!} \frac{d^n{\rm 
e}^{-\beta{\cal H}}}{d {\cal H}^n} : \underbrace{Q \cdot \cdots
 \cdot Q}_{n} 
\end{equation}
in quantum analysis. Thus the {\it n}-th order nonlinear response is
expressed by the {\it n}-th order quantum derivative of $\rho$.
The above expression (3$\cdot$1) of higher derivatives of $f({\cal H}) 
= {\rm e}^{-\beta {\cal H}} $ in 
terms of the inner derivation $\delta_{{\cal H}}$ is convenient for 
evaluating the required nonlinear responses explicitly, 
for example, using the 
high-temperature expansion method. The above static perturbational 
expansion with respect to the external field {\it H}
can be extended to that of the nonequilibrium density 
matrix $\rho(t)$ given by a solution of the von Neumann equation 
(4$\cdot$1). 

\section{Basic equations in nonequilibrium systems}

As is well known, the density matrix $\rho(t)$ in a nonequilibrium 
system 
satisfies the von Neumann equation 
\begin{equation}
i \hbar \frac{d}{dt}\rho(t)=[{\cal H}(t), \rho(t)]=\delta_{{\cal 
H}(t)} \rho(t) 
\end{equation}
for the time-dependent Hamiltonian ${\cal H}(t)$ of the relevant 
system. 

Now we attempt to find a solution of the exponential form 
\begin{equation}
\rho(t)={\rm e}^{-\eta(t)}.
\end{equation}
Concerning the ``entropy operator" $\eta(t)$, we have the following 
formula which was pointed out by Zubarev.$^{4)}$

\vspace*{0.3cm}
\noindent
{\bf Formula 2 :} The entropy operator $\eta(t)$ defined in Eq.~(4$\cdot$2) 
satisfies the equation 
\begin{equation}
i \hbar \frac{d \eta(t)}{dt}=[{\cal H}(t), \eta(t)]. 
\end{equation}

This is a simple example of the following general formula.$^{3)}$ 

\vspace*{0.3cm}
\noindent
{\bf Formula 3 :} Any operator-valued function $f(\rho(t))$ of the 
density matrix $\rho(t)$ satisfies the equation 
\begin{equation}
i \hbar \frac{d}{dt}f(\rho(t))=[{\cal H}(t), f(\rho(t))]. 
\end{equation}

It is instructive to give here a compact proof due to quantum 
analysis: 
\begin{eqnarray}
i \hbar \frac{d}{dt}f(\rho(t)) & = & i \hbar \frac{df(\rho(t))}{d\rho(t)} 
\frac{d \rho(t)}{dt}
= \frac{df(\rho(t))}{d\rho(t)} \delta_{{\cal H}(t)} 
\rho(t)\nonumber\\  & = & -\frac{df(\rho(t))}{d\rho(t)} 
\delta_{\rho(t)} {\cal H}(t) = -\delta_{f(\rho(t))} {\cal H}(t)
= [{\cal H}(t), f(\rho(t))].
\end{eqnarray}
Here we have used Eq.(2$\cdot$16). The above equation (4$\cdot$3) is our starting 
point for deriving the renormalized expansion scheme (5$\cdot$10). 

\section{General perturbation theory on the entropy
operator in nonequilibrium systems}

We formulate here a general perturbation expansion of the
entropy operator for the 
Hamiltonian ${\cal H}(t)$ taking the form 
\begin{equation}
{\cal H}(t)={\cal H}-AF(t),
\end{equation}
with a time-dependent external force $F(t)$ (as in Kubo's linear 
response theory$^{5),6)}$). Here $A$ denotes an operator conjugate 
to the external force $F(t)$. Now, we define the correction term 
$\eta'(t)$ in 
\begin{equation}
\eta(t)=\Phi+\beta{\cal H} + \eta'(t)
\end{equation}
for $\eta(t)=-\log \rho(t)$, where 
$\beta = 1/k_{\rm B} T$ and
$\Phi$ is a normalization 
constant such that 
\begin{equation}
{\rm e}^{\Phi}={\rm Tr}\; \; {\rm e}^{-\beta {\cal 
H}-\eta'(-\infty)}. 
\end{equation}
We then expand the correction term $\eta'(t)$ as 
\begin{equation}
\eta'(t)=\sum^{\infty}_{n=1} \eta_n(t),
\end{equation}
so that $\eta_n(t)$ is of $n$th order in $F(t)$. This is a new 
type of renormalized perturbation theory for nonequilibrium
systems, because even the first-order 
term $\eta_1(t)$ gives partially infinite-order terms in the 
density matrix $\rho(t)$. 
It is easily shown from Formula 2 that $\eta'(t)$ satisfies the 
inhomogeneous equation 
\begin{equation}
\frac{d}{dt} \eta'(t)=\frac{1}{i \hbar}[{\cal H}(t), \eta'(t)] - 
\beta F(t) \dot{A}
\end{equation}
with the initial condition $\eta'(-\infty)=0$, which corresponds 
to the condition 
\begin{equation}
\rho(-\infty)=\rho_{{\rm eq}}={\rm e}^{-\beta {\cal H}}/{\rm 
Tr}\;{\rm e}^{-\beta {\cal H}}. 
\end{equation}
Here we have also used the relation
\begin{equation}
\dot{A} =\frac{1}{i \hbar}[A,{\cal H}]=\frac{1}{i \hbar} \delta_A 
{\cal H}. 
\end{equation}
Equation (5$\cdot$5) is the basic formula derived here. A new 
aspect of this equation is that it has the 
temperature-dependent source term $-\beta F(t)\dot{A}$. Since ${\cal 
H}(t)$ contains an external force $F(t)$, Eq.~(5$\cdot$5) is nonlinear with 
respect to this force. The linearized equation is 
given by 
\begin{equation}
\frac{d}{dt} \eta_1(t)=\frac{1}{i \hbar}[{\cal H}, \eta_1(t)] -\beta 
F(t)\dot{A}. 
\end{equation}
The solution of Eq.~(5$\cdot$8) with the initial condition $\eta_1(-\infty)=0$ 
is obtained as 
\begin{equation}
\eta_1(t) = -\beta \int^t_{-\infty}F(s) {\rm 
exp}\left(\frac{1}{i\hbar}(t-s)\delta_{\cal H}\right) \dot{A}ds = 
-\beta \int^0_{-\infty}{\rm e}^{\varepsilon s}F(t+s) \dot{A}(s)ds. 
\end{equation}
The adiabatic factor ${\rm e}^{\varepsilon s}$ 
has been inserted to insure convergence.

The above first-order approximation $\{\rho_1(t)={\rm 
exp}[-\Phi-\beta{\cal H}- \eta_1(t)]\}$ gives Zubarev's statistical 
operator$^{4)}$ when the Hamiltonian ${\cal H}(t)$ is given by 
${\cal H}(t)={\cal H}-AF(t)$. This first-order approximation, namely 
Zubarev's theory, is justified if the second-order term $\eta_2(t)$ 
is much smaller than $\eta_1(t)$. 

For higher-order 
correction terms of $\eta'(t)$, we have the following. 

\vspace*{0.3cm}
\noindent
{\bf Formula 4 :} The higher-order entropy operators $\{\eta_n(t)\}$ 
are given by 
\begin{eqnarray}
\eta_2(t) & = & -\frac{\beta}{i \hbar} \int^0_{-\infty} {\rm 
e}^{\varepsilon s}ds F(t+s) \int^s_0F(t+s')[A(s'), 
\dot{A}(s)]ds'\nonumber\\ & & \cdots \cdots \cdots\nonumber\\
\eta_n(t) & = & -\frac{\beta}{(i \hbar)^{n-1}} \int^0_{-\infty}{\rm 
e}^{\varepsilon s}ds F(t+s) \int^s_0 dt_1 \int^{t_1}_0 dt_2 \cdots 
\int^{t_{n-2}}_0 dt_{n-1} \nonumber\\ 
& & \times F(t+t_1) \cdots F(t+t_{n-1}) \delta_{A(t_1)} 
\delta_{A(t_2)} \cdots \delta_{A(t_{n-1})} \dot{A}(s) 
\end{eqnarray}
with the hyperoperator $\delta_{A(t)}$ and with $\dot{A}=(i 
\hbar)^{-1} \delta_A{\cal H}$. 
 
These formulas can be derived from Eq.~(5$\cdot$5). They will be 
useful in studying nonlinear responses, because they are
renormalized perturbational expansions in contrast to the 
ordinary perturbational expansion$^{5)}$ of the density matrix 
itself. In fact, even the above $\rho_1(t)$ includes terms up to 
infinite order in $F(t)$. Thus, our formulation (5$\cdot$10) is a 
new useful result, compared with the ordinary expansion scheme 
of $\rho(t)$ itself. The quantum analysis of dissipative
systems$^{13),14)}$ will be presented in Appendix C, using ordered exponentials
and free Lie elements.$^{15),16)}$

\section{Linear response in terms of the inner derivation}

In this section we discuss linear response as an application
of the general perturbation theory presented in the preceding
section, and we express it in terms of the inner derivation 
$\delta_{\cal H}$ for the relevant Hamiltonian ${\cal H}$.

The density matrix $\rho (t)$ for the Hamiltonian ${\cal H}(t)$
in Eq~(5$\cdot$1) is given by
\begin{eqnarray}
\rho & = & {\rm e}^{-\Phi -(\beta {\cal H} + \eta_1 (t))}
\nonumber \\
& = & {\rm e}^{-\Phi} \biggl({\rm e}^{-\beta {\cal H}}
+ \frac{d {\rm e}^{-\beta {\cal H}}}{d(\beta {\cal H})}
\eta_1 (t)\biggr)
\end{eqnarray}
up to first-order of in external force $F(t)$.
Here, $\eta_1 (t)$ is given by Eq.~(5$\cdot$9). The quantum
derivative $d {\rm e}^{-\beta {\cal H}}/d(\beta {\cal H})$
is expressed by
\begin{equation}
\frac{d {\rm e}^{-\beta {\cal H}}}{d (\beta {\cal H})}
= -e^{-\beta {\cal H}} \Delta (\beta {\cal H}),
\end{equation}
as is seen from Eq.~(2$\cdot$6{\rm b}) .
Thus, the first-order term $\Delta \rho (t)$ 
is given by
\begin{equation}
\Delta \rho (t) = -{\rm e}^{-\Phi} {\rm e}^{-\beta {\cal H}}
\Delta (\beta {\cal H}) \eta_1 (t).
\end{equation}
The average of the relevant current operator
$J = \dot{A} $ is expressed as
\begin{eqnarray}
\langle J \rangle_t 
& = & {\rm Tr} \Delta \rho (t) 
 J  =  -\langle
(\Delta (\beta {\cal H}) \eta_1 (t)) 
 J \rangle
\nonumber \\
& = & \beta \int^0_{-\infty } 
{\rm e}^{\varepsilon s} F (t + s)
\langle \Delta(\beta {\cal H} ) J(s)) 
 J \rangle ds
\nonumber \\
& = & \beta \int^{\infty }_0 
{\rm e}^{-\varepsilon s} F(t-s)
\langle (\Delta(\beta {\cal H}) 
J (-s)) J \rangle ds
\nonumber \\
&  \equiv & {\rm Re} (\sigma (\omega)
Fe^{ i \omega t}) 
\end{eqnarray}
under the assumption that 
${\rm Tr} \, J \, {\rm exp}(-\beta {\cal H}) = 0$
and $F(t) = F \cos (\omega t)$.
Here $<\cdots>$ denotes the average with respect to the
equilibrium density matrix, and the general 
conductivity $\sigma(\omega)$ is expressed as
\begin{equation}
\sigma(\omega) = \beta \int^{\infty}_0
{\rm e}^{-\varepsilon s - i \omega s}
\langle (\Delta(\beta {\cal H}) J) J 
(s) \rangle ds;
\end{equation}
namely
\begin{eqnarray}
\sigma(\omega) & = & \beta \langle (\Delta
(\beta {\cal H}) J)
\frac{1}{ i \omega -( i / \hbar)
\delta_{\cal H}} J \rangle
\nonumber \\
& = & \frac{\beta}{ i \omega}
\sum^{\infty}_{n=0} \langle(\Delta
(\beta {\cal H}) J)
[ (\frac{1}{\hbar \omega} \delta_{\cal H}
)^n J ] \rangle
\end{eqnarray}
for the Planck constant $\hbar$, using
the hyperoperator $\Delta(A)$ defined by
Eq.~(2$\cdot$7{\rm a}).

It is also interesting to note that we have 
\begin{equation}
\int^{\beta}_0 {\rm e}^{\lambda {\cal H}} J {\rm e}^{-\lambda {\cal 
H}}d \lambda = \int^{\beta}_0 {\rm e}^{\lambda \delta_{{\cal H}}}d 
\lambda J= \beta \Delta(\beta {\cal H})J 
\end{equation}
for the current operator $J$ in our notation. This may be thoght of as the 
`dressed current operator', due to quantum fluctuation. 
Thus, Kubo's canonical correlation $\langle J : J(t) \rangle$ is 
expressed as 
\begin{equation}
\langle J : J(t) \rangle \equiv \frac{1}{\beta} \int^{\beta}_0 
\langle {\rm e}^{\lambda {\cal H}}J {\rm e}^{-\lambda {\cal H}} J(t) 
\rangle d \lambda 
= \langle (\Delta(\beta {\cal H}) J)J(t) \rangle .  
\end{equation}
Then, the Kubo formula for the 
frequency-dependent conductivity $\sigma(\omega)$ is expressed 
in the form 
\begin{equation}
\sigma(\omega) = \beta \int^{\infty}_0 \langle J : J(t) \rangle {\rm 
e}^{-i \omega t}dt = \beta \langle (\Delta(\beta {\cal H}) 
J)\frac{1}{i \omega -(i/\hbar)\delta_{{\cal H}}} J \rangle.  
\end{equation}
In particular, we obtain
\begin{equation}
\sigma(\omega) \simeq \frac{\beta}{i \omega} \langle (\Delta(\beta 
{\cal H}) J) J \rangle 
\end{equation}
for large $\omega$. Some remarks on applications of Eqs.~(6$\cdot$6) and 
(6$\cdot$9) will be given in the succeeding section.

The present derivation of the Kubo formula may be more transparent and
the algebraic structure that $\sigma(\omega)$ is expressed only in
terms of the commutators of ${\cal H}$ and $J$ (namely free Lie elements)
is convenient in practical calculations, as will be shown elsewhere.

\section{Some remarks on the conductivity $\sigma (\omega )$}

It is instructive to give some remarks on applications of the 
formulas (6$\cdot$6) and (6$\cdot$9) for the conductivity $\sigma(\omega)$.

When the current $J$ is a constant of motion, $\sigma(0)$ is 
infinite$^{17),18)}$ as seen from (6$\cdot$5). We consider the following 
more general situation that the current operator $J$ contains 
some (not necessarily all) constants of motion $\{H_j\}$, that is,
\begin{equation}
J= \sum_j a_jH_j +J',
\end{equation}
where $J'$ is defined by the remaining part of $J$ orthogonal
to all the $\{ H_j \}$; namely 
$J'$ is off-diagonal with respect to energy$^{18)}$ (i.e., 
$\langle m \mid J' \mid n \rangle = 0$ for $E_m=E_n$ with the energy 
eigenvalues $\{E_n\}$ of the Hamiltonian ${\cal H}$ even in a 
degenerate case). Here, the 
coefficients $\{a_j\}$ in (7$\cdot$1), namely the ergodicity constants,
 are given$^{18)}$ by
\begin{equation}
a_j=
\langle J H_j \rangle / \langle H^2_j \rangle, 
\end{equation}
using the orthogonality condition
\begin{equation}
\langle H_j H_k \rangle = \langle H^2_j \rangle 
\delta_{jk}.
\end{equation}
Thus, the zero frequency (or static isolated) 
conductivity defined by 
\begin{equation}
\sigma(0)= \beta \int^{\infty}_0
\langle J : J(t) \rangle dt
\end{equation}
is seen to diverge as
\begin{equation}
\sigma(0)= \beta \sum_j a^2_j \int^{\infty}_0 \langle H_j^2 
\rangle dt +({\rm finite}) \rightarrow \infty, 
\end{equation}
when at least one of the ergodicity constants
$\{a_j\}$ is non-vanishing.
This remark is useful in practical applications$^{6),19)}$
of the Kubo formula 
to some exactly soluble systems$^{17),18)}$ with an infinite number 
of constants of motion.

\section{Concluding remarks}

The quantum analysis introduced in previous papers$^{1)-3)}$ has
been extended to the case of an unbounded operator $A$ in a Hilbert space by 
restricting our consideration to the three typical operator functions 
${\rm e}^{-A}, 1/A$ and $\log A$ under the situation that the 
differential $dA$ is bounded. The proof is rather easy but it is 
instructive for studying more difficult cases for unbounded 
operators. 

Our new expressions of response functions in terms of the inner 
derivation $\delta_{{\cal H}}$ (or the dressed current operator 
$\Delta(\beta {\cal H})J$) are convenient for analytic and numerical 
calculations of these response functions. This result should be 
compared with the abstract operator representation of a KMS-state by 
Naudts, Verbeure and Weder$^{20)}$ in the more complicated situation 
of infinite systems. 

The renormalized perturbation scheme of the density matrix $\rho(t)$ 
is one of the new results in the present paper. This is in sharp
contrast to Kubo's well-known systematic expansion formula$^{5)}$ 
of $\rho(t)$ itself, rather than $\log \rho(t)$.  

It is also interesting to note that the quantum analysis is useful 
in expressing an exponential product of a dissipative density matrix
in terms of a single exponential 
(namely the generalized BCH formula) composed only of commutators, 
as is exemplified in Appendix C.

Transport coefficients are also expressed in terms of commutators of 
the relevant current operators. 

\section*{Acknowledgements}

The present author would like to thank Dr.~H.~L.~Richards,
Dr.~H.~Kobayashi, Dr.~G.~Su and Dr.~H.~Asakawa
for useful comments. This work has been supported
by the CREST (Core Research for Evolutional Science and
Technology) of the Japan Science and Technology Corporation
(JST). He would also like to thank Noriko Suzuki for
continual encouragement.

\appendix
\section{Convergence of Eq.~(2$\cdot$5)}

The convergence of Eq.~(2$\cdot$5) is shown as follows:
\begin{eqnarray}
\lefteqn{\mid \mid
({\rm e}^{-(A+hB)}-{\rm e}^{-A})/h
-\int^1_0 {\rm e}^{-(1-s)A}(-B) {\rm e}^{-sA} ds \mid 
\mid}\nonumber\\ & & \leq \mid \mid \frac{1}{h}
\int^h_0 dt \int^1_0 ds \left[{\rm e}^{-(1-s)(A+tB)}B \{ {\rm 
e}^{-s(A+tB)}-{\rm e}^{-sA}\}\right.\nonumber\\ & & \left.+ \{{\rm 
e}^{-(1-s)(A+tB)}-{\rm e}^{-(1-s)A}\} B {\rm e}^{-sA} \right] \mid 
\mid\nonumber\\ & & \leq \frac{\mid \mid h \mid \mid \cdot \mid \mid 
B \mid \mid^2}{2} \max_{\mid t \mid \leq \mid h \mid}\left[\int^1_0 
ds \mid \mid {\rm e}^{-(1-s)(A+tB)} \mid \mid \right.\nonumber\\ & & 
\times \int^s_0 d \lambda \mid \mid
{\rm e}^{-(s-\lambda)(A+tB)} \mid \mid \cdot \mid \mid {\rm 
e}^{-\lambda(A+tB)} \mid \mid\nonumber\\ & & \left.+ \int^1_0 ds 
\int^{1-s}_0 d\lambda \mid \mid {\rm e}^{-(1-s-\lambda)(A+tB)} \mid 
\mid \cdot \mid \mid {\rm e}^{-\lambda(A+tB)} {\rm e}^{-sA} \mid \mid 
\right],  
\end{eqnarray}
when $A$ is a positive (but unbounded) operator and $B=dA$ is 
bounded. Therefore, we arrive at Eq.~(2$\cdot$5).

\section{Expansion Formulas and Convergence of Higher-Order 
Derivatives}
 
The $n$th derivative of $f(A)$ is given by Eq~(3$\cdot$1), namely by
the following integral:$^{1)}$  
\begin{equation}
\frac{d^nf(A)}{dA^n}=n ! \int^1_0 dt_1 \cdots \int^{t_{n-1}}_0 dt_n 
f^{(n)}(A-\sum^n_j t_j\delta_j).
\end{equation}
Here, $\delta_j$ is a hyperoperator defined by Eq.~(3$\cdot$2), namely by 
\begin{equation}
\delta_j : (dA)^n=(dA)^{j-1} (\delta_A dA)(dA)^{n-j}.
\end{equation}
This is also formally expanded as follows.

\noindent
{\bf Formula A :}
\begin{equation}
\frac{d^nf(A)}{dA^n}=\sum^{\infty}_{m=0} \frac{n!(-1)^m}{m 
!}f^{(n+m)}(A) 
\int^1_0 dt_1 \cdots \int^{t_{n-1}}_0 dt_n (\sum^n_j t_j\delta_j)^m.
\end{equation}
For example, the first derivative $df(A)/dA$ is given by Eq.~(2$\cdot$17), and 
\begin{eqnarray}
\frac{d^2f(A)}{dA^2} & = &
\sum^{\infty}_{m=0} \frac{2!(-1)^m}{(m+2)!} f^{(m+2)}(A) 
\frac{1}{\delta_2}\left[(\delta_1+\delta_2)^{m+1} 
-\delta_1^{m+1}\right], \nonumber\\ \frac{d^3f(A)}{dA^3} & = &
\sum^{\infty}_{m=0} \frac{3!(-1)^m}{(m+3)!} f^{(m+3)}(A) 
\{\frac{\delta_1^{m+2}}{\delta_2(\delta_2+ \delta_3)}\nonumber\\ &- 
&\frac{(\delta_1+\delta_2)^{m+2}}{\delta_2\delta_3} + 
\frac{(\delta_1+\delta_2+\delta_3)^{m+2}}{(\delta_2+ 
\delta_3)\delta_3}\}, \cdots. \quad \quad \quad \quad \quad \quad 
\quad \quad \quad \quad \quad \hspace*{-0.2cm}  
\end{eqnarray}

The expansion of
\begin{equation}
d^nf(A) \equiv \frac{d^nf(A)}{dA^n} \; : \; (dA)^n 
\end{equation}
converges in the uniform norm topology when 
\begin{equation}
\alpha_n \equiv \lim^{\mbox{---}}_{m \rightarrow \infty} \mid \mid 
\frac{f^{(n+m)}(A)}{m!}
\int^1_0 dt_1 \cdots \int^{t_{n-1}}_0dt_n
(\sum^n_j t_j\delta_j)^m( dA)^n \mid 
\mid^{\frac{1}{m}} < 1.
\end{equation}
 
\section{Quantum Analysis of Dissipative Density Matrices} 

It is instructive to discuss first the  
non-dissipative unitary case.

\vspace*{0.3cm}
\noindent
(i) Unitary case. Here we discuss the von Neumann 
equation 

\begin{equation}
i \hbar \frac{d}{dt} \rho(t)=[{\cal H}(t), \rho(t)] =\delta_{{\cal 
H}(t)}\rho(t) 
\end{equation}
for the time-dependent Hamiltonian ${\cal H}(t)$ of the relevant 
system, as in (4$\cdot$1). A formal solution of Eq.~(C$\cdot$1) is given by 
\begin{eqnarray}
\rho(t) & = & {\rm exp}_{+}\left(\frac{1}{i \hbar} \int^t_0 
\delta_{{\cal H}(s)}ds\right) \rho(0)\nonumber\\ & = & {\rm 
exp}_{+}\left(\frac{1}{i \hbar} \int^t_0 {\cal H}(s)ds\right)\rho(0) 
{\rm exp}_{-}\left(-\frac{1}{i \hbar} \int^t_0 {\cal H}(s)ds\right). 
\end{eqnarray} 
Here, we have used the following ordered exponentials:$^{13),14)}$ 
\begin{equation}
{\rm exp}_{+} \int^t_0 A(s) ds=1+ \int^t_0 A(s)ds + \cdots + \int^t_0 
dt_1 \cdots \int^{t_{n-1}}_0 dt_n A(t_1) \cdots A(t_n)+ \cdots  
\end{equation}
and
\begin{equation}
{\rm exp}_{-} \int^t_0 A(s) ds=1+ \int^t_0 A(s)ds + \cdots + \int^t_0 
dt_1 \cdots \int^{t_{n-1}}_0 dt_n A(t_n) \cdots A(t_1)+ \cdots.  
\end{equation}
Clearly the ordered exponentials
\begin{equation}
{\rm exp}_{+}\left(\frac{1}{i \hbar} \int^t_0 {\cal H}(s)ds\right) \; 
\; {\rm and} \; \; {\rm exp}_{-}\left(-\frac{1}{i \hbar} \int^t_0 
{\cal H}(s)ds\right) 
\end{equation}
are both unitary and consequently are bounded when $\{{\cal H}(s)\}$ 
are self-adjoint. Thus, the arguments in \S~2 can also be 
applied to these ordered exponentials, namely operator 
functionals.$^{21)}$ It is then shown that the operator 
functional derivation$^{21)}$ for the variation $\delta{\cal H}(t_1)$ 
\begin{eqnarray}
dF[{\cal H}(t)]_{t_1} & \equiv &
d\left[{\rm exp}_{+}\left(\frac{1}{i \hbar} \int^t_0 {\cal 
H}(s)ds\right)\right]_{t_1} \equiv \frac{\delta F[{\cal 
H}(t)]}{\delta {\cal H}(t_1)} \cdot \delta{\cal H}(t_1)\nonumber\\ & 
= & {\rm exp}_{+}\left(\frac{1}{i \hbar} \int^t_{t_1} {\cal 
H}(s)ds\right) \delta{\cal H}(t_1) {\rm exp}_{+}\left(\frac{1}{i 
\hbar} \int^{t_1}_0 {\cal H}(s)ds\right) 
\end{eqnarray}
is bounded when the elements of $\{{\cal H}(t)\}$ are self-adjoint and 
the elements of $\{\delta 
{\cal H}(t_1)\}$ are bounded. Similarly the operator functional 
derivation of ${\rm exp}_{-}[-\frac{1}{i \hbar} \int^t_0 {\cal 
H}(s)ds]$ is bounded under the same conditions.

\vspace*{0.3cm}
\noindent
(ii) Dissipative case. We discuss here the unnormalized density 
operator $\hat{\rho}(t)$ of a dissipative system described by 
the master equation 
\begin{equation}
\frac{d \hat{\rho}(t)}{dt} =\frac{1}{i \hbar}[{\cal H}, 
\hat{\rho}(t)] + \Lambda \hat{\rho}(t) + \hat{\rho}(t) 
\Lambda^{\dagger}. 
\end{equation}
Here, $\Lambda$ and $\Lambda^{\dagger}$ denote some bounded 
operators expressing a dissipative effect. The Hamiltonian 
${\cal H}$ may be unbounded. The normalized density matrix 
$\rho(t)$ is given by $\rho(t)=N(t)\hat{\rho}(t)$ with 
$N(t)^{-1} = {\rm Tr} \hat{\rho}(t)$.
A formal solution of Eq.~(C$\cdot$7) is given as follows. First we put 
\begin{equation}
\hat{\rho}(t)={\rm exp}\left(\frac{t}{i \hbar} {\cal H}\right) f(t) 
{\rm exp}\left(-\frac{t}{i \hbar} {\cal H}\right). 
\end{equation}
Then, Eq.~(C$\cdot$7) can be rewritten as
\begin{equation}
\frac{df(t)}{dt}= \Lambda_t f(t) +f(t) \Lambda^{\dagger}_t, 
\end{equation}
where
\begin{equation}
\Lambda_t={\rm exp}\left(-\frac{t}{i \hbar} {\cal H}\right)\Lambda \; 
{\rm exp}\left(\frac{t}{i \hbar} {\cal H}\right)  
\end{equation}
and $\Lambda^{\dagger}_t = (\Lambda_t)^{\dagger}$. Next we put 
\begin{equation}
f(t)={\rm exp}_{+}\left(-\int^t_0 \Lambda^{\dagger}_s ds\right)g(t) 
{\rm exp}_{-}\left(\int^t_0 \Lambda^{\dagger}_s ds\right) 
\end{equation}
with $g(0)=f(0)=\hat{\rho}(0)$. Then Eq.~(C$\cdot$9) 
can again be rewritten as 
\begin{equation}
\frac{dg(t)}{dt}={\cal L}(t)g(t), 
\end{equation}
where
\begin{equation}
{\cal L}(t)={\rm exp}_{-}\left(\int^t_0 \Lambda^{\dagger}_s ds\right) 
(\Lambda_t + \Lambda^{\dagger}_t)
{\rm exp}_{+}\left(-\int^t_0 \Lambda^{\dagger}_s ds\right).  
\end{equation}
A solution of Eq.~(C$\cdot$12) is given by
\begin{equation}
g(t)={\rm exp}_{+}\left(\int^t_0 {\cal L}ds\right) g(0).  
\end{equation}
Thus we arrive at
\begin{equation}
\hat{\rho}(t)={\rm exp}_{+}\left(\int^t_0 {\cal L}(s,t)ds\right) 
\hat{\rho}(t,0), 
\end{equation}
where
\begin{eqnarray}
{\cal L}(s,t)&=&{\rm exp}\left(\frac{t}{i \hbar} {\cal H}\right) {\rm 
exp}_{+}\left(-\int^t_0 \Lambda^{\dagger}_s ds\right) {\cal 
L}(s)\nonumber\\ & \times &
{\rm exp}_{-}\left(\int^t_0 \Lambda^{\dagger}_s ds\right) {\rm 
exp}\left(-\frac{t}{i \hbar} {\cal H}\right), 
\end{eqnarray}
and
\begin{eqnarray}
\hat{\rho}(t,0)&=& {\rm exp}\left(\frac{t}{i \hbar} {\cal H}\right) 
{\rm exp}_{+}\left(-\int^t_0 \Lambda^{\dagger}_s 
ds\right)\hat{\rho}(0)\nonumber\\ & \times &
{\rm exp}_{-}\left(\int^t_0 \Lambda^{\dagger}_s ds\right) {\rm 
exp}\left(-\frac{t}{i \hbar} {\cal H}\right). 
\end{eqnarray}
When both $\hat{\rho}(0)$ and $\Lambda$ are bounded, 
$\hat{\rho}(t)$ is also bounded.

Now we put
\begin{equation}
\hat{\rho}(t,0)={\rm e}^{-\eta(t,0)}. 
\end{equation}
Then, we have
\begin{equation}
\hat{\rho}(t)={\rm exp}_{+}\left(\int^t_0{\cal L}(s,t)ds\right) {\rm 
exp}(-\eta(t,0)). 
\end{equation}
Our purpose here is to find the logarithm of Eq.~(C$\cdot$19). For this, we put 
\begin{equation}
{\rm exp}_{+}\left(\int^x_0{\cal L}(s,t)ds\right) {\rm 
exp}(-\eta(t,0))={\rm e}^{\Phi(x)}.  
\end{equation}
Clearly we have $\Phi(0)= -\eta(t,0)$. By differentiating Eq.~(C$\cdot$20) with 
respect to $x$, we obtain
\begin{equation}
{\rm e}^{\Phi(x)} \Delta (-\Phi(x)) \frac{d \Phi(x)}{dx} = {\cal 
L}(x,t) {\rm e}^{\Phi(x)}.  
\end{equation}
This is transformed into the equation 
\begin{eqnarray}
\frac{d \Phi(x)}{dx} & = & \Delta^{-1}(-\Phi(x)){\rm 
e}^{-\delta_{\Phi(x)}} {\cal L}(x,t)\nonumber\\
& =&
\Delta^{-1}(\Phi(x)){\cal L}(x,t)\nonumber\\ & = & 
\frac{\delta_{\Phi(x)}}{{\rm e}^{\delta_{\Phi(x)}}-1} {\cal L}(x,t)= 
\frac{\log {\rm e}^{\delta_{\Phi(x)}}}{{\rm e}^{\delta_{\Phi(x)}}-1} 
{\cal L}(x,t),  
\end{eqnarray}
using the identity $\delta_{\Phi}=\log {\rm e}^{\delta_{\Phi}}$. Then we 
can apply Eq.~(C$\cdot$20) to Eq.~(C$\cdot$21). 
Thus we finally arrive at the following formula. 

\vspace*{0.3cm}
\noindent
{\bf Formula B :} The entropy operator $\hat{\eta}(t)$ of the system 
described by Eq.(C.7) is expressed in the form 
\begin{eqnarray}
\lefteqn{\hat{\eta}(t)=-\Phi(t) \equiv -\log \hat{\rho} 
(t)}\nonumber\\ & & = {\eta}(t,0)-\int^t_0dx
\frac{\log [{\rm exp}_{+}(\int^x_0 \delta_{{\cal L}(s,t)}ds) {\rm 
exp}(-\delta_{\eta(t,0)})]} {{\rm exp}_{+}(\int^x_0 \delta_{{\cal 
L}(s,t)}ds) {\rm exp}(-\delta_{\eta(t,0)})-1} {\cal L}(x,t). 
\end{eqnarray}

The final expression (C$\cdot$23) is much more convenient than 
\begin{equation}
\hat{\eta}(t)=-\log \left[{\rm exp}_{+}\left(\int^t_0 {\cal 
L}(s,t)ds\right) 
{\rm exp}(-{\eta}(t,0))\right], 
\end{equation}
because Eq.~(C$\cdot$23) is expressed in terms of the commutators of $\{{\cal 
L}(s,t)\}$ and ${\eta}(t,0)$, namely free Lie elements$^{15)}$. 

The present formulation can be easily extended to the following more 
general dissipative system :
\begin{equation}
\frac{d \hat{\rho}(t)}{dt}=\frac{1}{i \hbar}[{\cal H}(t), 
\hat{\rho}(t)] + \Lambda(t) \hat{\rho}(t)+ \hat{\rho}(t) 
\Lambda^{\dagger}(t).  
\end{equation}
The expression (C$\cdot$23) is convenient for studying quantum effects$^{16)}$ 
in non-equilibrium systems.



\end{document}